\documentclass[11pt]{article}
\usepackage[T1]{fontenc}
\usepackage[utf8]{inputenc}
\usepackage[a4paper,margin=2.5cm]{geometry}
\usepackage{lmodern}
\usepackage{amsmath,amssymb,amsthm}
\usepackage{booktabs}
\usepackage{graphicx}
\graphicspath{{figures/}}
\usepackage{xcolor}
\usepackage{hyperref}
\usepackage[capitalise,noabbrev]{cleveref}
\usepackage{microtype}
\newtheorem{lemma}{Lemma}
\newtheorem{theorem}{Theorem}
\newtheorem{corollary}{Corollary}
\newtheorem{proposition}{Proposition}
\theoremstyle{definition}
\newtheorem{definition}{Definition}
\theoremstyle{remark}

\newcommand{\R}{\mathcal{R}}
\newcommand{\eps}{\varepsilon}
\newcommand{\code}[1]{\texttt{\let\_\codeus #1}}
\newcommand{\codeus}{\textunderscore\allowbreak}

\title{Alert-Preserving Lossy Compression of Monitoring Time Series}
\author{Bogdan Serykh\\
  Faculty of Computer Science, HSE University, Moscow, Russia\\
  \href{mailto:baserykh@edu.hse.ru}{\code{baserykh@edu.hse.ru}}\quad
  ORCID \href{https://orcid.org/0009-0007-0249-9959}{0009-0007-0249-9959}}
\date{November 2026\\[2pt]
  \small Preprint, \href{https://doi.org/10.5281/zenodo.24504704}{doi:10.5281/zenodo.24504704}}

\begin{document}
\maketitle

\begin{abstract}
Monitoring systems store large volumes of floating-point time series, and
most of these data are read only by alerting rules: threshold tests on
window aggregates with a hold duration. A lossy compressor with one error
bound for all samples either wastes space far from the thresholds or
changes alerts near them. We use the alert decisions as the correctness
criterion. For every rule, the reconstructed series must produce the same
sequence of alert states (\textsf{Inactive}, \textsf{Pending},
\textsf{Firing}) as the original, at any evaluation interval and phase.
For PromQL threshold rules over \code{last},
\code{avg/sum/min/max\_over\_time} and \code{rate} we derive a sound error
budget for each sample, computable in linear time. It is the smallest
value, over all windows that contain the sample, of the window's distance
to the threshold divided by a Lipschitz constant. Because the aggregates
are monotone, the budget splits into separate bounds for downward and
upward errors. For Prometheus' extrapolated \code{rate}, an
order-preserving quantizer leaves only the first and last sample of each
window constrained. Our codec combines predictive coding, a range coder and
a \emph{lazy} step choice that keeps the prediction whenever the budget
allows, which suits the asymmetric budget. After encoding, it decodes the
output, re-evaluates every rule and repairs any changed decision, so the
guarantee holds for the floating-point output. Parts of the
floating-point argument are machine-checked with Kani. On 13 real series
from the public Prometheus demo servers (7 and 30 days), the codec keeps
every alert decision (9\,055 firings) at 1.0--1.5 bits per value. This is
4--6.5 times less than a uniform bound tuned on the same data and less
than Compression Safeguards over SZ3 on 19 of 26 rule variants. In an
ablation, the one-sided budget and the lazy step choice together reduce
the size by a factor of 2.4--2.5.
\end{abstract}

\section{Introduction}
\label{sec:intro}

Monitoring systems such as Prometheus~\cite{prometheus} store large
volumes of floating-point time series (metrics). Few of the stored bytes
are ever read by a person. Most are read by alerting rules, which test a
window aggregate against a threshold, for example
\code{avg\_over\_time(x[5m]) > 0.8 for 10m} or
\code{rate(requests[5m]) < 20}.

Lossless time-series compressors of the Gorilla
family~\cite{pelkonen2015gorilla} spend 20--55 bits per value on typical
gauges, metrics such as load or memory use that move up and down
(\cref{def:kinds}). Error-bounded lossy compressors such as
SZ3~\cite{liang2023sz3} take a uniform pointwise bound $\eps$ and use it to
reduce the size. For alerting, one bound for all samples is a poor fit.
Far from a threshold almost any error is harmless. Near a threshold, an
error smaller than the measurement noise can create or suppress an alert.
A user who picks $\eps$ by hand therefore either wastes space or changes
which alerts fire without noticing.

We use the alert decisions themselves as the correctness criterion. A lossy
compressor is \emph{alert-preserving} for a rule set if every rule produces
the same alert state trajectory on the reconstruction as on the original,
that is, the same state (\code{Inactive}, \code{Pending} or \code{Firing})
at every evaluation. Any information the rules cannot observe can then be
discarded.

\paragraph{Contributions.}
\begin{itemize}
  \item A formal model of PromQL threshold alerts on a single series, with
    left-open windows, the \code{for}/\code{keep\_firing\_for} automaton
    and Prometheus' extrapolated \code{rate}. On this model we define
    decision preservation and \emph{phase robustness}, which requires
    preservation for every evaluation interval and phase without knowing
    how many samples a window holds (\cref{sec:model}).
  \item An error budget for each sample, derived from the rules
    (\cref{thm:budget}). An evaluation tolerates an error equal to its
    margin to the threshold divided by a Lipschitz constant of its
    aggregate, and a sample's budget is the smallest tolerance among the
    evaluations whose windows contain it. All aggregates except
    \code{rate} are monotone, so the budget splits into separate downward
    and upward bounds (\cref{cor:onesided}). For \code{rate} we show that
    counter resets make it discontinuous, and that under an
    order-preserving quantizer only the window ends need a budget
    (\cref{prop:rate}). The budget takes $O(n)$ time per rule.
  \item A codec that converts the asymmetric budget into a small stream
    (\cref{sec:codec}). It quantizes with closed-loop DPCM on power-of-two
    grids, chooses steps \emph{lazily} (it keeps the prediction when the
    budget allows), and codes the result with an adaptive binary range
    coder. A verify-and-repair loop compares the decoded output with a
    reference evaluator, so the guarantee holds in floating point. An
    optimistic variant starts from a multiple of the analytic budget and
    repairs only the evaluations that change, never going below the
    analytic budget.
  \item Bit-precise machine-checked proofs (Kani~\cite{vanhattum2022kani})
    of the floating-point budget for \code{last}, \code{min\_over\_time}
    and \code{max\_over\_time}, and of codec invariants
    (\cref{sec:verification}).
  \item An evaluation on synthetic series and on 13 real series from the
    public Prometheus demo servers with their upstream alert rules
    (\cref{sec:eval}). The 7- and 30-day snapshots are released with the
    code. We compare against uniform error bounds and against Compression
    Safeguards~\cite{tyree2026safeguards}, the closest existing system,
    and ablate every codec component.
\end{itemize}

\paragraph{Results.} On real data, no output of the rule-aware codec
changed any alert state, over 1\,864 firings in the 7-day snapshot and
7\,191 in the 30-day one. With thresholds that do fire, the codec needs
about 1--1.5 bits per value on real series. A uniform bound tuned by hand
on the same data needs 4--6.5 times more and still gives no guarantee on
other data. The codec is also smaller than Compression Safeguards over SZ3
on 19 of 26 real rule variants and on every synthetic one. In the
ablation, the one-sided budget and the lazy step choice together reduce
the size by a factor of 2.4--2.5.

The code and every table in this paper can be reproduced from
\url{https://github.com/ProgrammerPeasant/alert-preserving-compression}.
The real-series snapshots are archived at
\href{https://doi.org/10.5281/zenodo.24442344}{doi:10.5281/zenodo.24442344}~\cite{serykh2026snapshots}.

\section{Model and correctness criterion}
\label{sec:model}

We define series, compressors, rules and alert states, and then the
correctness criterion. The definitions do not assume knowledge of PromQL;
\cref{tab:promql} lists the Prometheus counterpart of each object.

\subsection{Series and compressors}

\begin{definition}[Series]
\label{def:series}
A \emph{series} of length $n$ is a finite sequence of samples
$X = ((s_i, x_i))_{i=1}^{n}$, where $s_i \in \mathbb{Z}$ is the timestamp
of sample $i$ in milliseconds, $s_1 < s_2 < \dots < s_n$, and
$x_i \in \mathbb{R}$ is its value. We write $I = \{1, \dots, n\}$ for the
index set.
\end{definition}

\begin{definition}[Value-only compressor]
\label{def:compressor}
A \emph{compressor} is a pair of maps: an encoder that turns a series into
a bit string and a decoder that turns the bit string back into a series. It
is \emph{value-only} if the decoded series has the same timestamps,
$Y = ((s_i, y_i))_{i=1}^{n}$; only the values may change. We write
$C(X) = Y$ for the composition of encoder and decoder and
$e_i = y_i - x_i$ for the pointwise error. $C$ is \emph{lossless} on $X$ if
$e_i = 0$ for all $i \in I$.
\end{definition}

Timestamps are compressed losslessly by existing delta-of-delta
schemes~\cite{pelkonen2015gorilla} and are out of scope; every compressor in
this paper is value-only.

Monitoring systems distinguish metric types by how their values may evolve.
Prometheus has gauges, counters, histograms and summaries; the classic
forms of the last two are stored as groups of gauge and counter
series~\cite{prometheus}, so two types suffice for our analysis.

\begin{definition}[Gauges and counters]
\label{def:kinds}
A series $X$ is a \emph{gauge} if its values are arbitrary reals: they
measure a current level that may go up and down (load average, memory in
use, temperature). It is a \emph{counter} if its values are non-negative
and count events cumulatively since the start of the monitored process
(requests served, bytes sent), so that they never decrease except at a
\emph{reset}, an index $i > 1$ with $x_i < x_{i-1}$, where the process
restarted and the count began again from zero.
\end{definition}

The kind is metadata of the series (the \code{TYPE} line of the exposition
format), not something inferred from the values. Counters are used through
their rate of increase (the aggregate $\mathrm{rate}$ below), gauges
through their level.

\subsection{Alerting rules}

\begin{definition}[Threshold rule]
\label{def:rule}
A \emph{threshold rule} is a tuple
\[
  R = (F,\ w,\ \bowtie,\ \theta,\ d_{\mathrm{hold}},\ d_{\mathrm{keep}}),
\]
whose components are:
\begin{enumerate}
  \item $F$, the \emph{aggregate}: a function that maps a finite non-empty
    sequence of samples to a real number. We consider
    $F \in \{\mathrm{last}, \mathrm{avg}, \mathrm{sum}, \mathrm{min},
    \mathrm{max}, \mathrm{rate}\}$. For values $v_1, \dots, v_m$,
    $\mathrm{last} = v_m$, $\mathrm{avg} = \frac1m \sum_j v_j$,
    $\mathrm{sum} = \sum_j v_j$, and $\mathrm{min}$, $\mathrm{max}$ are the
    extrema; $\mathrm{rate}$ is the per-second increase of a counter, which
    also depends on the timestamps (\cref{sec:budget-rate});
  \item $w > 0$, the \emph{range}: the length of the time window the
    aggregate reads, in milliseconds;
  \item ${\bowtie} \in \{>, \ge, <, \le\}$, the \emph{comparison};
  \item $\theta \in \mathbb{R}$, the \emph{threshold};
  \item $d_{\mathrm{hold}} \ge 0$, the \emph{hold duration}: how long the
    condition must stay true before the alert fires;
  \item $d_{\mathrm{keep}} \ge 0$, the \emph{keep duration}: how long a
    firing alert keeps firing after the condition has become false.
\end{enumerate}
\end{definition}

For example, the Prometheus alerting rule
\begin{center}
\code{expr: avg\_over\_time(node\_load1[5m]) > 2}\qquad \code{for: 10m}
\end{center}
is the tuple $R = (\mathrm{avg},\ 5\,\mathrm{min},\ {>},\ 2,\
10\,\mathrm{min},\ 0)$. \Cref{tab:promql} lists the correspondence for
every component.

\begin{table}[htbp]
\centering
\caption{Notation of \cref{sec:model} and its counterpart in a Prometheus
alerting rule.}
\smallskip
\label{tab:promql}
\begin{tabular}{@{}lll@{}}
\toprule
notation & Prometheus & meaning \\
\midrule
$X$ & one time series (a single label set) & the input \\
gauge, counter & metric type (\code{\# TYPE} line) & kind of $X$ \\
$F = \mathrm{last}$ & instant selector \code{x} & most recent sample \\
$F \in \{\mathrm{avg}, \mathrm{sum}, \mathrm{min}, \mathrm{max}\}$
  & \code{avg\_over\_time}, \code{sum\_over\_time}, \dots & window aggregate \\
$F = \mathrm{rate}$ & \code{rate} & counter increase per second \\
$w$ & range in \code{x[w]} & window length \\
$\bowtie$, $\theta$ & comparison and constant in \code{expr} & threshold test \\
$d_{\mathrm{hold}}$ & \code{for} & hold duration \\
$d_{\mathrm{keep}}$ & \code{keep\_firing\_for} & keep duration \\
$T$ & evaluation interval and phase & evaluation grid \\
$\lambda$ & lookback delta (5\,min) & staleness limit of \code{last} \\
\bottomrule
\end{tabular}
\end{table}

\paragraph{Evaluation.} Rules are evaluated at instants of an
\emph{evaluation grid} $T = (t_1 < t_2 < \dots)$; in Prometheus the grid is
fixed by the evaluation interval and an unknown phase. At an instant $t$ the
rule reads the samples whose index lies in its \emph{support}
\[
  S_R(t) = \{\, i \in I : t - w < s_i \le t \,\}.
\]
The window is open on the left, as range selectors are in Prometheus~3. For
$F = \mathrm{last}$ the expression has no range; the support is the single
latest sample within the lookback delta, $S_R(t) = \{\max\{ i : t - \lambda <
s_i \le t \}\}$ with $\lambda = 5$\,min, or $\emptyset$ if there is none.
Given the support, the rule computes two quantities:
\begin{itemize}
  \item the \emph{aggregate value}
    $f_R(X)(t) = F\bigl((s_i, x_i)_{i \in S_R(t)}\bigr)$, defined when
    $S_R(t) \ne \emptyset$;
  \item the \emph{condition}
    $c_R(X)(t) = \bigl[\, S_R(t) \ne \emptyset \ \wedge\ f_R(X)(t) \bowtie
    \theta \,\bigr] \in \{0, 1\}$.
\end{itemize}
An empty support makes the condition false, as Prometheus returns no sample
then.

\subsection{Alert states}

Whether an alert fires also depends on the hold and keep durations, so the
rule has to remember past conditions. Prometheus keeps this memory in a
small state machine (\code{rules/alerting.go}~\cite{promsrc}); we restate it as an automaton.

\begin{definition}[Alert automaton]
\label{def:automaton}
The automaton $M_R$ has states $Q = \{\mathsf{Inactive}, \mathsf{Pending},
\mathsf{Firing}\}$ and two timers, the activation time $a$ and the keep
start $b \in \mathbb{Z} \cup \{\bot\}$. It starts in $\mathsf{Inactive}$
with $b = \bot$ and reads pairs $(t_j, c_j)$, where $c_j = c_R(X)(t_j)$. At
each step $j = 1, 2, \dots$ it applies, in order:
\begin{enumerate}
  \item if $c_j = 1$: set $b \gets \bot$; if the state is $\mathsf{Inactive}$,
    move to $\mathsf{Pending}$ and set $a \gets t_j$;
  \item if $c_j = 0$: if the state is $\mathsf{Firing}$ and
    $d_{\mathrm{keep}} > 0$, set $b \gets t_j$ when $b = \bot$, and stay in
    $\mathsf{Firing}$ while $t_j - b < d_{\mathrm{keep}}$; otherwise move to
    $\mathsf{Inactive}$ and set $b \gets \bot$;
  \item if the state is $\mathsf{Pending}$ and
    $t_j - a \ge d_{\mathrm{hold}}$, move to $\mathsf{Firing}$.
\end{enumerate}
The state after step $j$ is $q_j$. The \emph{alert state trajectory} of $R$
on $X$ over $T$ is $D_R(X) = (q_1, q_2, \dots) \in Q^{|T|}$.
\end{definition}

With $d_{\mathrm{hold}} = 0$ step~3 fires on the first true condition, so
$\mathsf{Pending}$ is never observed; with $d_{\mathrm{keep}} = 0$ a false
condition always resets the alert.

\subsection{Correctness criterion}

\begin{definition}[Decision preservation]
\label{def:preservation}
Let $\R$ be a finite set of rules. A compressor $C$ is
\emph{decision-preserving} for $\R$ on $X$ over a grid $T$ if
$D_R(C(X)) = D_R(X)$ for every $R \in \R$. It is \emph{phase-robust} if it
is decision-preserving over every grid $T$.
\end{definition}

The criterion compares whole trajectories. It therefore also requires
the same $\mathsf{Pending}$ periods, which is stricter than requiring the
same firings. Relaxations (firing times within $\delta$, firings only)
admit larger budgets and are left for future work
(\cref{sec:conclusion}).

The two lemmas below reduce decision preservation to preserving the
condition at finitely many instants.

\begin{lemma}[Automaton]
\label{lem:automaton}
If $c_R(X)(t) = c_R(Y)(t)$ for all $t \in T$, then $D_R(X) = D_R(Y)$.
\end{lemma}
\begin{proof}
$M_R$ reads only the instants $t_j$ and the conditions $c_j$, which coincide
for $X$ and $Y$; its transitions are deterministic.
\end{proof}

\begin{lemma}[Phase robustness]
\label{lem:phase}
Let $F \ne \mathrm{rate}$, so that $f_R(X)(t)$ depends on $t$ only through
the support $S_R(t)$. Let
$E_R = \{s_i : i \in I\} \cup \{s_i + w : i \in I\}$ (with $\lambda$ in
place of $w$ for $F = \mathrm{last}$). Then $S_R(t)$, and hence $c_R(X)(t)$
and $c_R(Y)(t)$, is constant on every interval between consecutive points
of $E_R$, and $S_R(t) = \emptyset$ before the first point.
\end{lemma}
\begin{proof}
By definition $i \in S_R(t) \iff s_i \le t < s_i + w$, so the membership of
$i$ changes only at $t = s_i$ or $t = s_i + w$. For $\mathrm{last}$ the
support is the largest index in the same kind of window with $\lambda$
in place of $w$, which changes only at the same kind of points.
\end{proof}

Call the points of $P_R = E_R$ the \emph{protected instants} of $R$; there
are at most $2n$ of them. By \cref{lem:phase}, every instant $t$ has the
same condition as some protected instant (the last one not after $t$), for
both $X$ and $Y$. With \cref{lem:automaton} this gives:

\begin{corollary}
\label{cor:protected}
If $c_R(C(X))(t) = c_R(X)(t)$ for every $R \in \R$ with
$F \ne \mathrm{rate}$ and every $t \in P_R$, then $C$ is phase-robust for
these rules.
\end{corollary}

The encoder therefore needs the rule's range $w$. It does not need the
evaluation interval, the phase, the scrape interval or the number of
samples per window, so the guarantee also covers missed scrapes and
irregular timestamps. The
$\mathrm{rate}$ aggregate depends on $t$ continuously through its
extrapolation to the window edges (\cref{sec:budget-rate}); for it the
conditions are preserved on the deployed grid $T$ only, $P_R = T$.

\section{Error budget}
\label{sec:budget}

\subsection{A sufficient condition}

Consider a protected evaluation $(R,t)$ with a non-empty support. Its
\emph{margin} $m_R(t) = |f_R(X)(t) - \theta|$ is the distance of the
aggregate from the threshold. Let $L_R(t)$ be a Lipschitz constant of $F$
on $\mathbb{R}^{|S_R(t)|}$ in the sup norm. For \code{last},
\code{avg\_over\_time}, \code{min\_over\_time} and \code{max\_over\_time}
we have $L = 1$, and for \code{sum\_over\_time} $L = |S_R(t)|$. The
\emph{tolerance} of the evaluation is $\tau_R(t) = m_R(t)/L_R(t)$. A sample
is read by every evaluation whose support contains it, and its budget is
the smallest tolerance among them:
\begin{equation}
  \eps_i = \min\{\tau_R(t) : R \in \R,\ t \in P_R,\ i \in S_R(t)\},
  \qquad \min\emptyset = +\infty .
  \label{eq:budget}
\end{equation}

On the original data the condition is $f_R(X)(t) \bowtie \theta$ or its
negation, so it is one of $f > \theta$, $f \ge \theta$, $f < \theta$ and
$f \le \theta$. Two properties of this inequality are used below.

\begin{definition}[Tight, upper and lower evaluations]
\label{def:tight}
A protected evaluation $(R,t)$ is \emph{tight} if the inequality that holds
on $X$ is strict, i.e.\ $c_R(X)(t) = 1$ with ${\bowtie} \in \{>,<\}$, or
$c_R(X)(t) = 0$ with ${\bowtie} \in \{\ge,\le\}$. Then a reconstruction with
$f_R(Y)(t) = \theta$ already flips the condition, so the error must stay
strictly below the tolerance. It is \emph{upper} if $f_R(X)(t)$ lies on
the side of $\theta$ that only a decrease of the aggregate can leave
($c_R(X)(t) = 1$ with ${\bowtie} \in \{>,\ge\}$, or $c_R(X)(t) = 0$ with
${\bowtie} \in \{<,\le\}$), and \emph{lower} otherwise.
\end{definition}

The \emph{box budget} is the set of reconstructions
$\{Y : |y_i - x_i| \le \eps_i \text{ for all } i \in I\}$. It bounds the
error of each sample by the same amount in both directions.

\begin{theorem}[Box budget]
\label{thm:budget}
If $|y_i - x_i| \le \eps_i$ for all $i$, strictly on the supports of tight
evaluations, then $c_R(Y)(t) = c_R(X)(t)$ for all
$R \in \R$ and $t \in P_R$, and hence $D_R(Y) = D_R(X)$ on every grid (on $T$
for \code{rate}).
\end{theorem}
\begin{proof}
Supports depend on timestamps only. By the Lipschitz bound
$|f_R(Y)(t) - f_R(X)(t)| \le L_R(t)\max_{i\in S_R(t)}|e_i| \le m_R(t)$,
strictly when tight, so $f_R(Y)(t)$ stays on the same side of $\theta$.
\Cref{lem:phase,lem:automaton} finish the proof.
\end{proof}

The Lipschitz step is standard quantity-of-interest error
propagation~\cite{jiao2022qoi}. Our setting differs in where errors and
margins live: errors belong to samples, margins belong to evaluations, and
each sample is read by every evaluation whose window contains it. The
budget is therefore a minimum over overlapping windows, which has two
consequences. First, an encoder must look ahead by the longest window
$\max_R w_R$ before it can fix a sample's budget, so it fits block
compaction (Prometheus compacts 2-hour blocks) or a buffering proxy better
than the write path. Second, a sample with margin zero must be stored
exactly. Any smaller budget is also sufficient, so a pointwise cap
$\min(\eps_i, \eps_{\max})$ can bound the error for dashboards without
weakening the alert guarantee.

\subsection{One-sided budget}

Near a threshold a series usually stays on one side of it, and only an
error towards $\theta$ can change a decision. The box budget of
\cref{thm:budget} also limits errors away from $\theta$, which is not
necessary.

\begin{corollary}[One-sided budget]
\label{cor:onesided}
Let $\delta^-_i$ (the \emph{downward} budget) be the minimum of
$\tau_R(t)$ over the upper evaluations whose support contains $i$, and
$\delta^+_i$ (the \emph{upward} budget) the same minimum over the lower
ones, with $\min\emptyset = +\infty$ (\cref{def:tight}). If
$-\delta^-_i \le y_i - x_i \le \delta^+_i$ for all $i$ (strict where
tight), every protected condition is preserved.
\end{corollary}
\begin{proof}
All aggregates above are non-decreasing in each argument. For an upper
$(R,t)$ let $z_i = \min(y_i, x_i)$ on its support. Then
$|z_i - x_i| \le \delta^-_i \le \tau_R(t)$, so $z$ satisfies
\cref{thm:budget} at $(R,t)$, and $F(y) \ge F(z)$ keeps $f_R(Y)(t)$ on the
upper side. Lower evaluations are symmetric with $\max$.
\end{proof}

\begin{figure}[t]
\centering
\includegraphics[width=\linewidth]{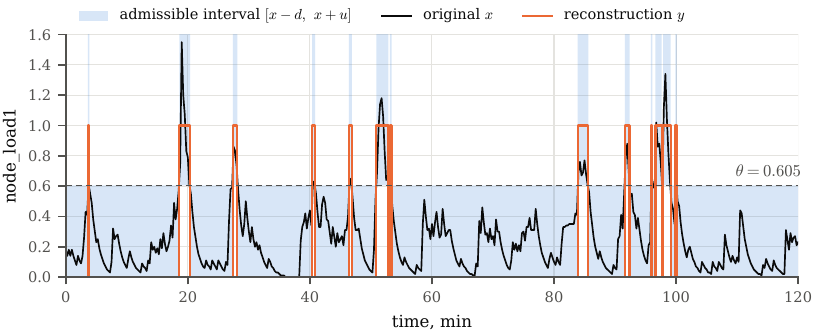}
\caption{One-sided budget and the codec's reconstruction on two hours of
\code{node\_load1} (7-day snapshot, rule \code{last > }$\theta$ with $\theta$
at the 95th percentile). Below $\theta$ only the upward error is bounded,
and above $\theta$ only the downward error. The admissible interval
(shaded) is therefore a half-line, and the reconstruction takes only two
levels. No alert decision changes.}
\label{fig:budget}
\end{figure}

Since $\eps_i = \min(\delta^-_i, \delta^+_i)$, every reconstruction in the
box budget also satisfies the one-sided bounds. The one-sided set is much
larger: a sample read only by upper evaluations may move up by any amount
(\cref{fig:budget}). The codec can use this freedom only if it places
reconstructions asymmetrically (\cref{sec:codec}). With round-to-nearest
quantization, the one-sided budget alone gains little
(\cref{sec:ablation}).

\subsection{Counters and \code{rate}}
\label{sec:budget-rate}

We restate Prometheus' \code{rate} (\code{extrapolatedRate} in
\code{promql/functions.go}~\cite{promsrc}) for one evaluation $(R,t)$ with
$|S_R(t)| \ge 2$. Let $f$ and $l$ be the first and last indices of the
support $S_R(t)$. The rule reads:
\begin{itemize}
  \item the \emph{increase}
    $\Delta = x_l - x_f + \rho$, where
    $\rho = \sum_{i \in S_R(t),\, i > f,\, x_i < x_{i-1}} x_{i-1}$ adds back
    the value lost at every reset inside the window;
  \item the \emph{sampled interval} $\sigma = s_l - s_f$, the time covered
    by the samples;
  \item the \emph{extrapolation distances} $h_{\mathrm{s}}$ and
    $h_{\mathrm{e}}$ from the first and last sample to the window's start
    $t - w$ and end $t$. Each is replaced by half the mean sample spacing
    when it reaches $1.1$ times that spacing. Both depend on timestamps
    only;
  \item the \emph{zero distance} $h_0 = \sigma x_f / \Delta$, used only
    when $\Delta > 0$. It is how long before $s_f$ the counter was zero if
    it grew at the window's average speed.
\end{itemize}
The increase is extrapolated to the window boundaries, but not past the
counter's zero point, and divided by the range in seconds:
\[
  \mathrm{rate} = \frac{\Delta}{w}\cdot
  \frac{\sigma + \min(h_{\mathrm{s}}, h_0) + h_{\mathrm{e}}}{\sigma}.
\]

\begin{proposition}
\label{prop:rate}
\leavevmode
\begin{enumerate}
  \item[(a)] The reset indicator $[x_i < x_{i-1}]$ compares two samples. An
    arbitrarily small error can therefore create a reset and change
    $\Delta$ by a whole counter value, and in general no positive
    symmetric budget preserves \code{rate}.
  \item[(b)] The closed-loop floor quantizer
    $y_i = y_{i-1} + q\lfloor (x_i - y_{i-1})/q\rfloor$, which stores the
    sample exactly when the step count would be negative, gives
    $x_i - q < y_i \le x_i$ and preserves
    $x_i \ge x_{i-1} \Rightarrow y_i \ge y_{i-1}$. If, in addition, the two
    samples around each true reset are stored exactly, the set of resets
    is preserved.
  \item[(c)] On reconstructions that preserve the resets, $\mathrm{rate}$ is
    Lipschitz in $(\Delta, x_f)$ away from $x_f = 0$. For a densely
    sampled window with $r$ resets the constant is about $(2 + r)/w$.
  \item[(d)] Under (b), $\Delta$ and $x_f$ depend only on the first and last
    samples of the window and on the preserved reset values. The budget
    of a \code{rate} evaluation therefore applies only to the two ends of
    its window.
\end{enumerate}
\end{proposition}

The implementation does not use the constant from (c). It computes a
sound interval enclosure of the \code{rate} formula, including the
zero-point branch, over the box $y_i \in [x_i - \eps, x_i]$, and finds the
largest safe $\eps$ by bisection. With a 30\,s evaluation grid and 15\,s
scrapes, about half of a counter's samples are not the end of any window.
By (d), only monotonicity constrains them.

\subsection{Where no budget exists}

A positive budget exists at $x$ if and only if $x$ lies in the interior of
the set of inputs that give the same condition. Some expressions break this
at their decision points: equality tests, \code{changes()}, \code{floor}
and \code{round} inside an expression, and aggregations that select series
by label, such as \code{topk}. They need exact samples at those points.
Ratios and \code{histogram\_quantile} are locally Lipschitz and can be
handled with an interval enclosure, as for \code{rate}.

\subsection{Floating point and complexity}

The budget is computed in floating point, and the code subtracts a
rounding slack from the margin. A floating-point sum of $n$ terms has error
at most $\gamma_{n-1}\sum_i|v_i|$, with $\gamma_n = nu/(1 - nu)$ and unit
roundoff $u = 2^{-53}$~\cite{higham2002accuracy}. For strict comparisons
the code takes the next representable value below the tolerance.
Monotonicity holds bitwise for \code{last}, \code{min} and \code{max}.
Prometheus computes sums with Kahan compensation, which is not monotone, so
for \code{sum/avg} the proof of \cref{cor:onesided} uses the error bound
instead. An evaluation whose margin is within the rounding slack is
constrained in both directions.

Supports move monotonically with $t$, so two pointers find all of them. A
two-stack sliding queue computes the window aggregates, and a monotone
deque computes the per-sample minimum in \eqref{eq:budget} as a sliding
minimum over instants. The budget costs $O(n + |P_R|)$ per rule. For
\code{rate}, each instant adds a bisection of about a dozen enclosure
evaluations.

\section{Codec}
\label{sec:codec}

The codec takes the per-sample bounds $(\delta^-_i, \delta^+_i)$ and
produces a bit stream whose decoded values lie in the \emph{admissible
interval}
\[
  A_i = [\,x_i - \delta^-_i,\ x_i + \delta^+_i\,].
\]
A budget of zero (or NaN) forces the sample to be stored exactly, and a
budget of $+\infty$ leaves it unconstrained. The stream format does not
depend on the rules, and the decoder reads only the stream.

\paragraph{Quantizer.} The lossy mode is closed-loop DPCM (differential
pulse-code modulation) on a grid of step $q = 2^e$. The prediction for
sample $i$ is the previous \emph{reconstruction}, $p_i = y_{i-1}$. Because
the prediction does not use the original values, the encoder and the
decoder compute the same prediction and errors do not accumulate. The
reconstruction is $y_i = p_i + k_i q$, where the integer \emph{step count}
$k_i$ is chosen from the admissible counts
\[
  K_i = \{\, k \in \mathbb{Z} : p_i + kq \in A_i \,\}.
\]
This set is non-empty whenever $q \le \delta^-_i + \delta^+_i$. The stream
stores either $k_i$ or the difference $k_i - k_{i-1}$.

\paragraph{Blocks.} The series is split into blocks of 128 samples, and
each block is coded either losslessly with Gorilla XOR
coding~\cite{pelkonen2015gorilla} or with the quantizer above, using one
exponent $e$ for the whole block. The encoder picks whichever is cheaper.
Each sample of the block proposes a candidate exponent: the largest $e$
with $2^e \le \delta^-_i + \delta^+_i$, so that $K_i \ne \emptyset$ (for a
counter, $2^e \le \delta^-_i$). The encoder also tries the three next
finer grids, because on discrete data a smaller step can give a smaller
stream. For a counter the quantizer is the floor quantizer of
\cref{prop:rate}(b), which keeps the reconstruction below the data and
non-decreasing.

A sample is \emph{escaped}, that is, stored as a raw 64-bit value, if its
step count would exceed $2^{50}$ in magnitude or if its reconstruction
misses $A_i$ in floating point. This also covers samples whose interval is
too narrow for the block's grid. The encoder checks every reconstructed
value against its bounds, computed bit for bit as the decoder will compute
it, so the bounds hold for the actual floating-point output.

\paragraph{Lazy steps.} The \emph{nearest} choice
$k_i = \mathrm{round}((x_i - p_i)/q)$ spreads errors evenly in both
directions. This suits a symmetric bound, but wastes most of an asymmetric
one. For every grid the encoder therefore also tries the \emph{lazy}
choice, the $k \in K_i$ closest to zero (so $y_i = p_i$ whenever the
prediction is admissible) or closest to $k_{i-1}$. With one-sided bounds
the prediction often stays admissible for long runs of samples, and the
entropy coder codes such runs at a few hundredths of a bit per sample. Per
block, the encoder compares four strategies (nearest or lazy, combined
with coding counts or differences) by their exact coded size.

\paragraph{Entropy coding.} Every block decision and every integer is
coded with an LZMA-style adaptive binary range coder~\cite{martin1979range}
with 15-bit probabilities. An integer (a step count or a difference) is
sent as a zero flag, a sign and its bit length in unary. These symbols are
modelled in a context given by the magnitude of the previous integer.
Then come the bit just below the leading one, which is also modelled, and
the remaining bits, which are written raw. The models adapt quickly at the
start of a stream and more slowly as their counts grow, so short streams
also compress well. For the ablation we keep a fixed-width variant that
stores the integers of a block with the block's largest bit length.

\paragraph{Chunks.} Prometheus cuts head chunks at 120 samples, and each
chunk must decode on its own. The codec can code the series as
independent streams, at 3--4 bytes of overhead per stream. The budget is
still computed on the whole series, because windows cross chunk
boundaries. A counter chunk starts from the previous chunk's last
reconstruction as a floor. Otherwise the output could drop at a chunk
boundary, and \code{rate} would see a reset that is not in the data.

\paragraph{Verify and repair.} After encoding, the codec decodes the
stream and re-evaluates every protected condition with the reference
semantics. For every evaluation whose condition changed, it sets the
budget of the evaluation's support to zero and encodes the series again.
The loop terminates, because a support stored exactly gives the original
values. For \code{sum/avg}, three components sum in different orders: the
verifier uses a sliding sum, the reference evaluator sums left to right,
and Prometheus uses Kahan summation. Their results differ by at most
$2\gamma_{n-1}\sum|y_i|$ (\cref{sec:budget}). The verifier therefore
accepts an evaluation only if its aggregate is farther than this from
$\theta$ or its support is bitwise exact. The analytic budget keeps twice
this margin and has not needed a repair on any dataset. The loop makes the
guarantee hold end to end with respect to the reference evaluator.

\paragraph{Optimistic budget.} The bounds of \cref{cor:onesided} assume
that all errors take their worst values at once, while in a window
aggregate the actual errors mostly cancel. The encoder can therefore start
from $\kappa\delta^\pm_i$ with $\kappa > 1$. On every evaluation that
changes, it halves the budget of the support, stopping at the analytic
budget, and sets it to zero once it has reached the analytic budget. By
\cref{thm:budget}, a support within the analytic budget cannot change its
evaluation. The analytic budget is therefore both the sound starting point
for $\kappa = 1$ and the floor that bounds the number of rounds. Compression
Safeguards~\cite{tyree2026safeguards} correct violations in a similar way,
but our encoder corrects only the supports whose evaluations changed.

\section{Machine-checked proofs}
\label{sec:verification}

\Cref{thm:budget} is a statement about real numbers, while the code
computes margins, tolerances and aggregates in IEEE-754 double precision.
We check the floating-point implementation with
Kani~\cite{vanhattum2022kani}, a bounded model checker for Rust built on
CBMC~\cite{kroening2014cbmc} that encodes floating-point operations
bit-precisely. Each harness quantifies over all finite inputs of a fixed
shape, and over NaN and infinities where stated. The times below are
wall-clock times for Kani~0.68 (CBMC with CaDiCaL) on a 24-core x86-64
machine running 12 harnesses at once.

We first tried a single harness stating that the budget implies the same
condition, over a window of three samples. It did not finish in 6 hours for
any aggregate. Symbolic execution is fast, but the final SAT query over the
chain of floating-point operations (aggregate, subtraction,
\code{next\_down}, comparison) is too hard. We therefore prove
\cref{thm:budget} for \code{last}, \code{min\_over\_time} and
\code{max\_over\_time} in three parts. An abstract tolerance $\tau$
separates the expensive tolerance computation from the window structure:
\begin{enumerate}
  \item[(A)] \emph{Scalar tolerance}, one harness per comparison: if
    $\mathrm{fl}|y' - v| \le \tau(v)$, where $\tau$ is the implemented
    tolerance function, then $y'$ and $v$ are on the same side of $\theta$
    (56--69 min each).
  \item[(B)] \emph{Window lemmas} on at most four variables. The extremum
    is attained by some sample. For the arg-extrema $j$ and $k$,
    $\mathrm{fl}|y_j - x_k| \le
    \max(\mathrm{fl}|y_j - x_j|, \mathrm{fl}|y_k - x_k|)$ (198 and
    226 min). Rounded subtraction is monotone (130 min).
  \item[(C)] The implemented tolerance of a window equals $\tau(f(x))$
    bitwise ($\le 3$ min), and the aggregates are monotone
    (\cref{cor:onesided}; $\le 8$ s).
\end{enumerate}
Together, (A)--(C) give the floating-point version of \cref{thm:budget}
for these aggregates and every window length.

For the codec, we prove that the decoder reconstructs exactly the values
the encoder computed, bit for bit, for every block parameter and including
$-0.0$. We also prove the gauge and counter invariants of the quantizer.
The quantizer tries several candidate step counts (rounding, the lazy
choice, and one step inwards) and returns only a candidate
that passes a single acceptance check. A harness over the full candidate
search did not finish in 6 hours, so we prove the invariants for the
acceptance check applied to an arbitrary candidate (7 and 15 min, run
separately on a 16-core machine). Finally, we prove that the zigzag and
varint codes and the bit-level specification of the bit I/O round-trip.

\paragraph{Not yet proved.} \code{sum/avg\_over\_time} need a lemma on the
error of a rounded sum in place of (B). The range coder and the concrete
bit writer run out of memory or time in symbolic execution, because of
unbounded \code{Vec} growth and a loop bound that depends on a symbolic
counter. Property tests cover these parts. Round-trip tests run both
formats with NaN, infinite and zero budgets and with counters, and
$5 \times 2000$ random decision-preservation cases are checked against a
reference evaluator that recomputes every instant. The tests catch a
mutation that multiplies the budget by 3.

\section{Evaluation}
\label{sec:eval}

\subsection{Setup}

\paragraph{Real data.} We use 13 series of raw 15\,s scrapes without gaps
from the public Prometheus demo servers~\cite{promdemo}. They include
node\_exporter gauges (load, available memory, free space on the root
filesystem, clock offset), counters and gauges of Prometheus itself, a
Caddy request counter and counters of a demo service. There are two
snapshots, of 7 days (40\,320 samples per series) and 30 days (172\,800
samples). The servers keep only 31 days of data, so we release the
snapshots with the code.

The demo server runs the upstream node-mixin and prometheus-mixin
alerts~\cite{nodemixin}. Nine of them read a single series once constant
denominators are folded into $\theta$. Counters without an upstream rule
get the rules \code{rate(x[5m]) > $\theta$ for 5m} and
\code{rate(x[5m]) < $\theta$ for 10m}. On a healthy host the deployed
thresholds (\code{prod}) never fire. We therefore also run every series
with two recalibrated thresholds, computed from the rule's own aggregate:
the 95th percentile (\code{q95}, an alert that fires often) and the median
(\code{q50}, where the aggregate crosses $\theta$ all the time). Each such
threshold is placed midway between two distinct values, so that discrete
metrics never sit exactly on $\theta$. This gives 26 rule variants per
snapshot, with 1\,864 firings over 7 days and 7\,191 over 30 days.

\paragraph{Synthetic data.} Five seeded generators produce 7 days of
data at 15\,s: CPU utilisation, request latency, memory, a request
counter, and the same counter with 8\,\% missed scrapes, outages of
1--6\,min and traffic that stays close to the threshold. Each dataset has
two rules that fire a few times per day.

\paragraph{Evaluation.} Rules are evaluated every 30\,s, 7\,s after the
scrapes. We compare full alert state trajectories, computed by the
reference evaluator in the repository, which implements Prometheus~3
semantics. Sizes are in bits per value of the value stream, without
timestamps; an uncompressed value takes 64 bits.

\paragraph{Baselines.} Our own configurations all use the same codec and
range coder and differ only in the error budget. \emph{Lossless} is the
codec with a zero budget, which yields Gorilla XOR blocks.
\emph{Uniform-safe} is the largest uniform bound that satisfies
\cref{thm:budget}. \emph{Uniform-tuned} is found by sweeping uniform
bounds over powers of four of the value range: it is the largest bound
such that neither it nor any smaller bound in the sweep changes a decision
\emph{on these data}. A user tuning a bound by hand could do no better,
and the bound guarantees nothing on other data. \emph{SZ3+SG} is
SZ3~\cite{liang2023sz3} wrapped in Compression
Safeguards~\cite{tyree2026safeguards} (1.0.0rc3). It has one stencil
quantity of interest for every window length the rule can read at any
phase, with the comparison itself as the quantity and bound zero. We sweep
13 SZ3 bounds and report the smallest output among those that change no
decision. \emph{Rule-aware} is the analytic budget (\cref{cor:onesided}),
and \emph{$\kappa$} is the optimistic budget with $\kappa = 1024$.

\subsection{Real series}
\label{sec:eval-real}

\begin{table}[t]
\centering
\caption{Mean bits per value over the 13 real series. Zero changed decisions
in every rule-aware cell. ``/120'': Prometheus-sized independent chunks.}
\label{tab:real}
\begin{tabular}{llrrrrrrr}
\toprule
 & rules & lossless & unif.-safe & unif.-tuned & rule-aware & $\kappa$ & rule-aware/120 & $\kappa$/120 \\
\midrule
7\,d  & prod & 27.71 & 0.08 & 0.09 & 0.07 & 0.07 & 0.45 & 0.42 \\
      & q95  & 25.52 & 6.91 & 7.29 & 1.13 & 1.00 & 1.77 & 1.59 \\
      & q50  & 25.52 & 7.69 & 6.01 & 1.49 & 1.38 & 2.16 & 2.00 \\
\midrule
30\,d & prod & 27.77 & 0.08 & 0.08 & 0.06 & 0.06 & 0.44 & 0.41 \\
      & q95  & 25.59 & 7.99 & 5.72 & 0.96 & 0.86 & 1.58 & 1.43 \\
      & q50  & 25.59 & 7.47 & 5.91 & 1.15 & 1.08 & 1.80 & 1.68 \\
\bottomrule
\end{tabular}
\end{table}

\Cref{tab:real} summarises the real data. With the deployed thresholds
nothing fires, and every codec, including the uniform ones, needs less
than 0.1 bits per value, because a threshold that the data never approach
constrains nothing. This is the usual situation in production, but it
does not distinguish the methods. With the recalibrated thresholds, the
rule-aware codec is smaller than the tuned uniform bound by a factor of
6.5 (q95) and 4.0 (q50) on 7 days, and 6.0 and 5.1 on 30 days. Unlike the
tuned bound, it also guarantees the decisions. Per series, its size ranges
from 0.01 bits per value for free filesystem space, which rarely changes,
to 3--4.5 for dense counters under \code{rate} with the median threshold.
In Prometheus-sized chunks of 120 samples, the size grows by 0.6--0.7 bits
per value, mostly because of the fixed overhead per chunk and because the
models start each chunk untrained.

\subsection{Comparison with Compression Safeguards}
\label{sec:eval-sg}

\begin{figure}[t]
\centering
\includegraphics[width=\linewidth]{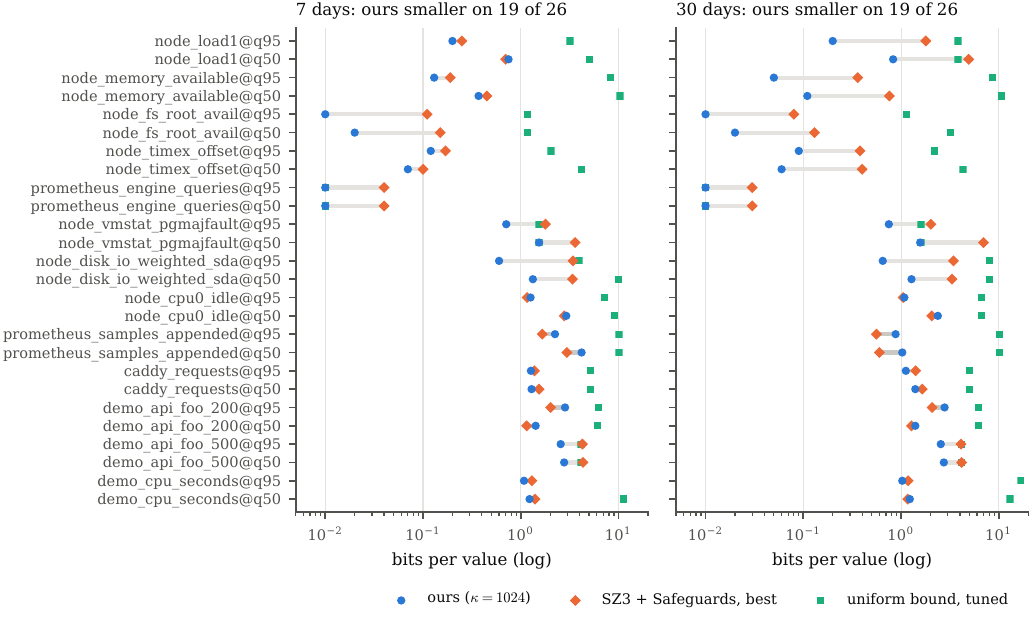}
\caption{Bits per value on the real series: our optimistic codec
($\kappa = 1024$), the smallest SZ3+SG output with no changed decision over
its bound sweep, and our codec with the largest uniform bound that keeps
every decision on these data (tuned, no guarantee). A grey bar joins ours
to SZ3+SG.}
\label{fig:sg}
\end{figure}

The optimistic codec is smaller than SZ3+SG on 19 of 26 variants in both
snapshots (\cref{fig:sg}). The geometric mean of the size ratio
$\kappa$/SG is 0.60 on 7 days and 0.42 on 30 days. SG stays smaller on
dense counters under \code{rate}
(\code{prometheus\_samples\_appended}, \code{node\_cpu0\_idle} and
\code{demo\_api\_foo\_200}). On these series our quantizer uses one step
for a whole 128-sample block, while SG corrects individual samples.

SG's translation of \code{rate} cannot express Prometheus' extrapolation.
It changed decisions at some sweep points on 4 of the 8 real counters
across the two snapshots, and at 11 of 13 sweep points on the synthetic
counter with missed scrapes. On that counter, the smallest SG output that
keeps every decision takes 8.45 bits per value, 2.9 times more than our
analytic codec (2.91) and 3.1 times more than our optimistic codec (2.74).
A user cannot tell in advance which SG bound is safe. SG needs up to 60\,s per sweep
point on a 30-day counter, while our whole verified codec runs at
11--16\,MB/s (\cref{sec:throughput}).

\subsection{Synthetic series}

\begin{table}[t]
\centering
\caption{Synthetic series, 7 days, bits per value (one stream per series).
SZ3+SG: smallest output with zero changed decisions over its sweep.}
\label{tab:synth}
\begin{tabular}{lrrrrrr}
\toprule
dataset & lossless & unif.-safe & unif.-tuned & SZ3+SG & rule-aware & $\kappa$ \\
\midrule
cpu            & 53.96 & 11.57 & 2.55 & 0.16 & 0.07 & 0.05 \\
latency        & 55.24 &  7.22 & 2.81 & 0.20 & 0.08 & 0.07 \\
memory         & 49.91 & 12.96 & 6.92 & 0.57 & 0.21 & 0.14 \\
requests       & 19.68 &  2.02 & 0.78 & 0.69 & 0.35 & 0.35 \\
requests-gappy & 21.76 & 11.46 & 9.46 & 8.45 & 2.91 & 2.74 \\
\bottomrule
\end{tabular}
\end{table}

\begin{figure}[t]
\centering
\includegraphics[width=0.5\linewidth]{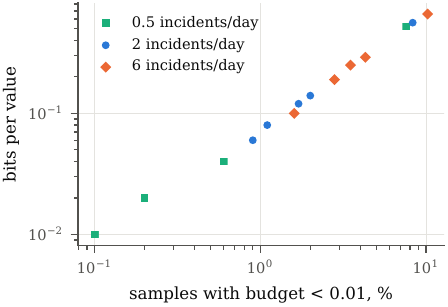}
\caption{Size of the rule-aware codec on synthetic \code{cpu} (7 days,
one rule) against the share of samples whose budget is below 0.01, for
$\theta \in \{0.55, \dots, 0.95\}$ and three incident rates.}
\label{fig:h2}
\end{figure}

On the synthetic gauges (\cref{tab:synth}) the rule-aware codec is
33--36 times smaller than the tuned uniform bound, and on the counters
2.2 and 3.2 times smaller. The analytic codec is smaller than SZ3+SG on
all five datasets. With the box budget of \cref{thm:budget} it was larger
than SZ3+SG on the three gauges, and this result led us to
\cref{cor:onesided}. The size grows with the share of samples near a
threshold: on \code{cpu}, it goes from 0.01 to 0.66 bits per value as that
share goes from 0.1\,\% to 10\,\% (\cref{fig:h2}).

\paragraph{Fidelity.} Without a cap, the reconstruction is useful only
for the rules, since a sample that is free in one direction drifts until
a bound stops it. With a pointwise cap $\eps_{\max}$, the rule-aware
codec is within 2\,\% of the size of a uniform bound $\eps_{\max}$ and
keeps every decision. A uniform bound of $10^{-3}$ of the range changes
17, 929 and 10\,655 alert states on \code{memory}, \code{requests} and
\code{requests-gappy}. Even the tuned uniform bound changes decisions on
two datasets when it is applied to 120-sample chunks instead of the
whole-series encoding it was tuned on. A bound tuned on one encoding of
the data does not carry over to another encoding of the same data.

\subsection{Ablation}
\label{sec:ablation}

\begin{table}[t]
\centering
\caption{Ablation: geometric mean of the size ratio to the full analytic
codec (row 3) over the synthetic series and the q95/q50 real variants
(7-day snapshot; the 30-day snapshot, in the repository, shows the same
ordering). Every cell keeps every
decision.}
\label{tab:ablation}
\begin{tabular}{lrrr}
\toprule
configuration & synthetic & real q95 & real q50 \\
\midrule
box budget, nearest steps (\cref{thm:budget}) & 2.44 & 2.53 & 2.47 \\
\quad + one-sided budget (\cref{cor:onesided}) & 2.27 & 2.06 & 1.91 \\
\quad + lazy steps (full analytic codec) & 1.00 & 1.00 & 1.00 \\
\quad + optimistic budget, $\kappa = 1024$ & 0.80 & 0.92 & 0.95 \\
full analytic + budget-scale search & 1.00 & 1.00 & 1.00 \\
full analytic $-$ one-sided budget & 1.37 & 1.21 & 1.20 \\
full analytic $-$ lazy steps & 2.27 & 2.06 & 1.91 \\
$\kappa = 1024$ $-$ one-sided budget & 0.93 & 1.05 & 1.08 \\
full analytic, fixed-width codes & 8.01 & 7.37 & 5.92 \\
\bottomrule
\end{tabular}
\end{table}

\Cref{tab:ablation} adds one component at a time. With the same stream
format, the one-sided budget and the lazy step choice reduce the size by
a factor of 2.4--2.5 when used together. Either one alone gains much
less. With round-to-nearest, the grid is chosen from the width of the
admissible interval and rounding uses that width symmetrically, so
one-sided bounds alone gain little and sometimes make the output larger.
On counters, all of the gain comes from lazy steps, which keep the
prediction on samples that are not the end of any window
(\cref{prop:rate}(d)). We had also encoded scaled copies of the budget
($\times\frac12, \frac14, \frac18$), because the coded size is not
monotone in the budget. Once the encoder tries the three next finer grids
per block, this search is redundant: it gives a smaller output on none of
the 31 datasets, so the codec no longer does it. The optimistic budget
saves a further 4--20\,\%, but on real data it does not replace the
one-sided budget. Entropy coding avoids the minimum of one bit per value
of fixed-width codes and makes the output 5.9--8.6 times smaller over
both snapshots.

\subsection{Throughput}
\label{sec:throughput}

We measure throughput on one core, in MB/s of raw float64 input, on
synthetic \code{cpu} and \code{requests} with 40\,320 samples. The $O(n)$
budget computation runs at 42--67\,MB/s. The encoder runs at 22\,MB/s,
since every block tries four grids with four step strategies each, and
the decoder at 2.5--2.9\,GB/s. The whole verified compressor runs at
11--16\,MB/s, or 3--4\,MB/s with the budget-scale search, which no longer
changes the size (\cref{sec:ablation}). A 2-hour Prometheus block holds
480 samples per series at 15\,s, so one core compresses about 2\,900
series-blocks per second.

\section{Related work}
\label{sec:related}

\paragraph{Correction-based safeguards.} Compression
Safeguards~\cite{tyree2026safeguards} wrap any compressor with declarative
requirements, such as pointwise bounds, sign preservation relative to an
offset, or bounds on quantities of interest over stencils. After
compression they compute corrections that enforce the requirements. A
window-average threshold test on a regular grid can be written as a stencil
quantity of interest: the quantity is the comparison itself and its bound
is zero. We run Safeguards this way as a baseline (\cref{sec:eval-sg}).

The two approaches differ in what has to be translated into the
requirement language and in how the requirement is enforced. Our
criterion needs no translation of hold durations (\cref{lem:automaton}),
but a stencil does need one for phase robustness. A stencil fixes the
window length in samples, so missed scrapes and an unknown evaluation
phase require one stencil for every window length that occurs, plus
truncated windows at the edges. Prometheus' \code{rate} extrapolation and
counter resets (\cref{prop:rate}) have no stencil form at all. The
translation we use for the baseline, the reset-corrected increase over the
sampled interval, changes decisions on counters with gaps. Safeguards
correct every point that violates a requirement. Our analytic budget is
sound from the start and needs no correction, and our optimistic variant
corrects only the supports whose evaluations changed, never going below
the analytic budget.

\paragraph{Quantity-of-interest preservation.} Jiao et
al.~\cite{jiao2022qoi} derive pointwise error bounds from bounds on derived
quantities by error propagation, and our Lipschitz step is one instance of
their method. We bound a decision instead of the error of a quantity, so
our budget depends on the distance to a threshold. We also handle the
structure in time: overlapping windows, evaluation phase and the alert
automaton.

\paragraph{Feature- and topology-preserving compression.} TopoSZ and its
successors~\cite{yan2024toposz} preserve critical points and contour trees
of scientific fields, and adaptive schemes tighten the bounds near
detected features~\cite{adaptivefeature2024}. Preserving the level set at
$\theta$ means preserving the sign of $x - \theta$, which is our setting
restricted to \code{last}. Window aggregates and alert states have no
counterpart in that work.

\paragraph{Time-series compression.} Gorilla~\cite{pelkonen2015gorilla} and
ALP~\cite{afroozeh2024alp} are lossless. SZ3~\cite{liang2023sz3} and recent
error-bounded time-series compressors~\cite{shrink2025} guarantee a uniform
pointwise bound. CAMEO~\cite{cameo2025} bounds the deviation of
autocorrelation functions, and Muñiz-Cuza et
al.~\cite{munizcuza2024forecasting} measure how error-bounded compression
affects forecasting. Plato~\cite{lin2020plato} gives deterministic error
bounds on aggregate queries over compressed series, which bound the
numeric result of a query. None of these systems bounds which side of a
threshold a result falls on, so none of them preserves discrete decisions.

\paragraph{Downstream guarantees and industrial practice.} Outside time
series, the closest guarantee we know is cluster preservation for particle
data~\cite{ren2026clusters}. Friends-of-friends cluster membership depends
on threshold tests on distances, and a correction applied after
compression, within the pointwise bound, restores it, much as our repair
loop restores changed evaluations. Industrial cost controls choose which
series to keep instead of bounding their values. Grafana's Adaptive
Metrics recommends aggregating away labels that no dashboard, rule or
recent query uses, and makes no recommendation for a metric that a
recording or alerting rule reads~\cite{grafana2026adaptive}. Our method
targets those metrics. We searched the literature (Crossref, arXiv,
EDBT/VLDB proceedings, 2015--2026) and vendor documentation and found no
compressor that guarantees alert decisions.

\section{Limitations and future work}
\label{sec:conclusion}

\paragraph{Scope of the rules.} We handle rules that read one series.
Aggregations across series (\code{sum by}) couple the budgets of all
series in a group, so the encoder would have to see the group together.
Ratios need an interval enclosure, and equality tests and \code{topk}
admit no positive budget at their decision points.

\paragraph{Known rules.} The budget is computed for a fixed rule set. A
rule added later reads a reconstruction that was optimised for other
rules, and only the optional pointwise cap bounds its error. The series
can be compressed again from the original as long as the original is kept,
for example until compaction.

\paragraph{Lookahead.} A sample's budget depends on evaluations up to one
window after it, so the encoder must buffer the longest window. This works
during compaction or in a buffering proxy, but not on a write path that
must add no latency.

\paragraph{Data.} Our real data are 13 series from two public demo
servers. The deployed thresholds never fire there, so we recalibrated the
thresholds to quantiles. An evaluation on production traces with more
rules, more incidents and gaps in the data is still needed.

\paragraph{Relaxations.} Requiring the same \textsf{Pending} periods makes
the criterion strict. If only firings had to be preserved, or firing times
up to $\delta$, the budget could exceed the margin on short condition
flips, since \code{for} filters them out. The analysis would then have to
reason about the automaton instead of single evaluations. The minimum in
\eqref{eq:budget} gives a feasible budget but not an optimal one. For
linear aggregates, the optimal split of a margin among the samples of a
window is a knapsack-like problem.

\section{Conclusion}

With alert decisions as the correctness criterion, lossy compression of
monitoring data has a sound per-sample error budget that can be computed
in linear time. The aggregates are monotone, so the budget is one-sided.
A coder that keeps its prediction whenever it is still admissible turns
this into an output 2.4--2.5 times smaller than with the symmetric
Lipschitz box. On real Prometheus series the codec needs from a few
tenths of a bit to a few bits per value, and no alert decision changes.

\bibliographystyle{plainurl}
\bibliography{refs}

\end{document}